\documentclass[10pt, conference, letterpaper]{IEEEtran} 

\usepackage{cite}
\usepackage{times}

\usepackage{latexsym}
\usepackage{amsfonts}
\usepackage{paralist}
\usepackage{setspace}
\usepackage{color}

\usepackage{url} 
\usepackage{graphicx, xspace}
\usepackage{epstopdf}   
\usepackage{subfigure}
\usepackage{caption}
\usepackage{indentfirst}
\usepackage{amsmath}
\usepackage[noend]{algorithmic}
\usepackage{algorithm}
\usepackage{amssymb}
\usepackage{multirow}
\usepackage[scaled=0.85]{beramono}

\newtheorem{theorem}{Theorem}

\newtheorem{definition}{Definition}

\makeatletter

\newcommand{\Rmnum}[1]{\expandafter\@slowromancap\romannumeral #1@}

\makeatother

\allowdisplaybreaks 
\def\ie{\textit{i.e.}\xspace}
\def\etal{\textit{et al.}\xspace}

\def\eg{\textit{e.g.}\xspace}

\def\spk{speaker recognition\xspace}
\def\Spk{Speaker recognition\xspace}
\def\sr{speech recognition\xspace}
\def\Sr{Speech recognition\xspace}
\def\sp{service provider\xspace}

\def\vc{voice conversion\xspace}

\def\vs{voice sanitizer\xspace}
\def\Vs{Voice sanitizer\xspace}

\def\Ks{Keyword spotting\xspace}
\def\ks{keyword spotting\xspace}
\def\VoiceMask{VoiceMask\xspace}

\newcommand{\CUTXY}[1]{{}}
\newcommand{\jnl}[1] {}

\newcommand{\jnote}[1]{{$\langle${\textbf{\textcolor{blue}{#1}}}$\rangle$}}

\renewcommand{\paragraph}[1]{\smallskip \noindent {\textbf{#1}}}
\graphicspath{{./}{../Figures/}{./Figures/}}
\begin{document}


\title{VoiceMask: Anonymize and Sanitize Voice Input on Mobile Devices} 

\author{
	\IEEEauthorblockN{Jianwei~Qian\IEEEauthorrefmark{1},
	Haohua~Du\IEEEauthorrefmark{1},
	Jiahui~Hou\IEEEauthorrefmark{1},
	Linlin~Chen\IEEEauthorrefmark{1},
	Taeho~Jung\IEEEauthorrefmark{2},
	Xiang-Yang~Li\IEEEauthorrefmark{3},
	Yu~Wang\IEEEauthorrefmark{4},
	Yanbo~Deng\IEEEauthorrefmark{1}
	}
	\IEEEauthorblockA{
	\IEEEauthorrefmark{1} Department of Computer Science, Illinois Institute of Technology\\
	\IEEEauthorrefmark{2} Department of Computer Science and Engineering, University of Notre Dame\\
	\IEEEauthorrefmark{3} School of Computer Science and Technology, University of Science and Technology of China\\
	\IEEEauthorrefmark{4} School of Information and Communication Engineering, University of North Carolina at Charlotte\\
	}
}

\maketitle
\thispagestyle{plain}
\pagestyle{plain}


\begin{abstract}
Voice input has been tremendously improving the user experience of mobile devices by 
 freeing our hands from typing on the small screen. 
Speech recognition is the key technology that powers voice input, and it is usually outsourced 
 to the cloud for the best performance.
However, the cloud might compromise users' privacy by identifying their identities by voice, learning their 
 sensitive input content via speech recognition, and then profiling the mobile users based on the content.
In this paper, we design an intermediate between users and the cloud, named \emph{\VoiceMask}, 
 to sanitize users' voice data before sending it to the cloud for speech recognition.
We analyze the potential privacy risks and aim to protect users' identities and sensitive input content from being disclosed to the cloud.
\VoiceMask adopts a carefully designed voice conversion mechanism that is resistant to several attacks.
Meanwhile, it utilizes an evolution-based keyword substitution technique to sanitize the voice input content.
The two sanitization phases are all performed in the resource-limited mobile device while still maintaining the 
 usability and accuracy of the cloud-supported speech recognition service.
We implement the \vs on Android systems and present extensive experimental results that validate the effectiveness and efficiency of our app.
It is demonstrated that we are able to reduce the chance of a user's voice being identified from 50 people by $84\%$ 
while keeping the drop of speech recognition accuracy within $14.2\%$.
\end{abstract}

\begin{IEEEkeywords}
Privacy protection, voice input, voice privacy, voice sanitizing, speech processing.
\end{IEEEkeywords}


\section{Introduction}
\label{sec:introduction}

Featuring hand-free communication, voice input has been widely applied in keyboard apps 
 (\eg Google, Microsoft, Sougou, and iFlytek keyboards), voice search (\eg Microsoft Bing, Google Search) 
 and artificial intelligence personal assistants (\eg Apple's Siri, and Amazon Echo) 
 on a range of mobile devices.
Voice input can greatly ease our lives by freeing our hands and attention from the time-consuming work of typing 
 on the small screen of mobile devices.
It is also one of the major means of human-computer communication for people who are visually impaired. 
The key technology that powers voice input is speech recognition, also known as speech-to-text conversion, where the vocal input
 of spoken language is recognized and translated into texts.
Due to the resource limitation on mobile devices, the speech recognition  is usually outsourced to the cloud 
 server for higher accuracy and efficiency \cite{king2003server,chang2011csr}, which are two key elements leading to a good user experience for voice input systems.

However, most of the existing voice input service providers 
 collect their users' speech records\footnote{This was 
 validated through recent news \cite{apple-news1,google-policy} and private communication with several researchers working in the voice input industry.}.
Because the voice datasets, once correlated with the identities of the speakers, will reveal various private information describing the individuals, we argue that the Personally Identifiable Information (PII) should not be disclosed to the cloud (\textit{e.g.,} MAC addresses/Device IDs/User IDs associated with the datasets). This can be achieved by leveraging existing anonymous networks such as Tor \cite{tor,I2P}. However, with such a na\"{i}ve sanitization voice records can still be de-anonymized by linkage attacks with the \emph{speaker recognition} technology. Linkage attacks are possible because different individuals have different voice characteristics (\textit{i.e.,} voice biometrics),
and such voice biometrics can serve as fingerprints of the speech datasets.

In this paper, our \textbf{goal} is to protect the speech data privacy for voice input users while maintaining the user experience. Our strategy, in general, is to perturb the speech data in order to remove the personally identifiable voice fingerprints for the sake of complete anonymity for individual users. Our research also benefits existing voice input service providers by providing them with better data security. Because users' privacy is preserved at the data level and no individuals will be identified from the datasets the companies possess, they are less likely to be held responsible for the privacy leak caused by data breach once it occurs.

\begin{figure}[t]
  \centering
  \includegraphics[width=0.4\textwidth]{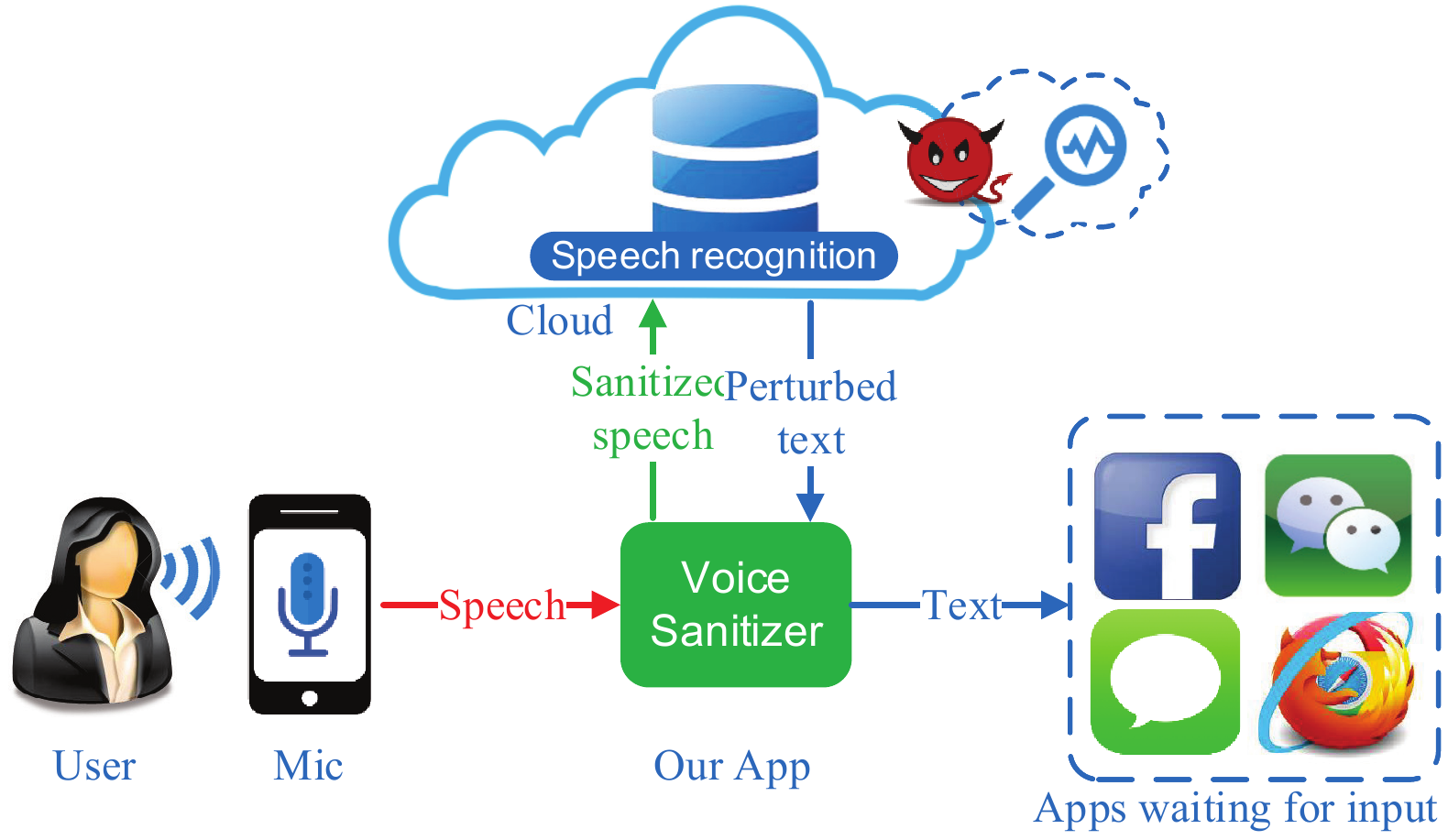}
  \caption{\scriptsize{\textbf{Our system architecture.} The cloud is a specialized \sr service provider. We design a secure voice input app on mobile devices that sanitizes the recorded voice input before sending it to the cloud. }}
  \label{scenario2}
\end{figure}

We are facing two \textbf{privacy risks} in the cloud-based voice input services. On one hand, the cloud
 (or third parties who obtained the voice data from the cloud) may link the speech records to individuals in real life
by harnessing the speaker recognition technology. If the cloud is able to collect some speech 
samples of the target person from other sources like YouTube and train a voice model of this person, 
then it can identify the records belonging to this person from the speech database, which leads to the user's \emph{identity privacy breach}.  
On the other hand, the cloud can also analyze the content of users' speeches and learn more detailed private information.
The cloud may use natural language processing techniques to extract information from a user's voice search history, 
voice command history, and even SMS messages and emails if they were typed via voice input, which leads to the user's \emph{content privacy breach}.
Then the cloud can paint a very accurate picture of the user's demographic categories, personal preferences, interpersonal communications, habits, schedules, and travels.
Therefore, the risk of privacy disclosure from speech data is real, and we believe it is very necessary to come up with countermeasures and
stop the leak of privacy from its source, that is, our smart mobile devices.
 
There are also serious \textbf{security risks} of existing voice input service architecture, such as \emph{spoofing attack} to speaker authentication systems. 
Voice as a type of biometric information has been widely used in emerging authentication systems to unlock smart devices, 
gain access to some apps like WeChat, and activate services like  Apple's ``Hey Siri'' and Google's ``OK Google''.
Once the cloud collects enough voice samples from a user,
it  can create an accurate voice print and use it to synthesize other speeches that sound like this user \cite{de2012evaluation,wu2014voice}, 
which may be exploited to spoof the voice authentication systems and gain access to the user's private resources. 
Worse still, the leaked voice can be used for fraud or to produce illegal recordings to frame and blackmail the person \cite{fakeObama,scam}.

In order to mitigate the aforementioned two privacy risks and avoid the security risk, we have two corresponding sub-goals in privacy preservation: 
1) preserving the users' identity privacy and 2) preserving their speech content privacy. 
The \textbf{challenges}  are fourfold.
Firstly, we have to sanitize users' speech data without degrading the accuracy of speech-to-text conversion to an unacceptable extent.
In other words, the users can still use the voice input app to transform their voice into text with high accuracy as before.
We should ensure the accuracy of \emph{speaker recognition} 
on the sanitized data by the cloud is greatly decreased while the accuracy of \emph{speech recognition} on them is well preserved. 
Secondly, it is challenging to protect \emph{speech content} privacy because it requires us to solve three sub-questions:
what information is considered sensitive/private? how to locate the sensitive content within the speech? 
how to hide the sensitive content without influencing the user experience?
Thirdly, speech as a type of unstructured data is hard to be processed in a privacy-preserving manner. Traditional secure multi-party computation methods 
cannot be employed directly. Cryptographic tools like searchable encryption for information retrieval do not apply to \sr either.
Lastly, it is hard to perform speech sanitization in real-time efficiently with the restricted resources on the mobile device.
The computation overhead should be small enough to induce acceptable latency for the voice input apps.
The power consumption  by the sanitization process should also be limited.

The \textit{na\"{i}ve method} is to simply perform speech recognition locally without the help from the cloud.
No privacy is leaked in this way, but the performance of offline \sr is inferior 
due to the limited computing resources on mobile devices \cite{mcgraw2016personalized}.
The local speech models cannot be updated timely as well. 
On the other hand, the cloud already provides free services so they tend to collect data as much as possible. Most of these companies are unwilling to release
offline \sr products. Google Nexus allows us to use offline voice typing but only when we are disconnected from the Internet.
As a result, we cannot expect the cloud to provide unconditional offline \sr service for users and we still have to rely on the powerful cloud to support more accurate speech recognition.
To protect privacy, we need to sanitize the speech data on the mobile device before sending it to the cloud for \sr. 
 

\textbf{Our solution:} 
To ensure the cloud has access to only the sanitized speech data, our basic solution is to introduce an intermediary app
(referred to as ``\VoiceMask'' hereafter) to perturb the speeches.
It acts as a third-party voice input app that connects users' mobile devices to the speech recognition cloud.
The \vs processes the audio signal received from the microphone and then sends the 
sanitized speech audio to the cloud. 
The sanitization is conducted from two aspects: (1) It disguises the speaker's identity by randomly 
modifying the speaker's voice using the voice conversion technology, and (2) it perturbs the contents of voice input via keyword substitution to hide the sensitive
information from being disclosed to the cloud. 
The two complementary perturbations performed locally in real time on the speech data ensure that the cloud (and the voice input app)  learns neither users' identity nor their private input contents.


\textbf{Contributions:} 
To the best of our knowledge, \VoiceMask is the first privacy-preserving architecture for the voice input function on mobile devices. 
We decouple the voice input app and the cloud to prevent the cloud from learning users' private information through the app (Section \ref{sec:solution}).
We then propose two light-weighted techniques for mobile devices to protect user identity and voice input content from being leaked to cloud servers.
The user identity privacy is preserved by a voice conversion technique that is resistant to the reversing attack or the reducing attack (Section \ref{identity}). 
The input content privacy is preserved by a randomized keyword substitution heuristic,
consisting of an evolution-based keyword spotting method for sensitive word detection and a differentially private algorithm for safeword set construction
(Section \ref{content}).
We implement \VoiceMask on Android systems and conduct extensive evaluation (Section \ref{sec:evalution}). 


\section{Background}
\label{sec:problem}
\subsection{System and Adversary Models}
The voice input service provider usually consists of an app on the user end 
and the cloud that performs speech recognition.
We assume it is honest but curious. 
It honestly executes the pre-designed protocol and provides a good user experience of voice input, 
but it might exploit the data it can access to learn the user's sensitive information. 
We also assume the data collected by the cloud is anonymous. For example, Apple has applied privacy-enhancing techniques to prevent their data analysts from seeing the user accounts associated with the data\footnote{Apple's privacy policy on data analytics, https://www.apple.com/privacy/approach-to-privacy/}.
Meanwhile, users who wish to achieve full anonymity can employ the anonymous network to stop the cloud from obtaining their PII. Please note this paper aims to defend against only voice input apps (and their cloud). There are certainly many other apps undermining our privacy, like browsers and emails, but how to defend against them is not our focus.

\subsection{Privacy Risks}
Most voice input service providers collect their users' speech data.
For instance, Apple stores our Siri voice commands for 24 months \cite{apple-news1} and Google stores everything we say to Google Now by default \cite{google-policy}. Thought they claim they will not sell our data in the privacy policy, they do analyze our data, which may compromise our privacy.
There are two types of privacy in this paper's scope: identity privacy and content privacy.

To  attack the \textit{identity privacy} of a target person,
the \sp may gather the person's speech audios from another source, for example from her YouTube channel or her posts on online social media. Then, a voice model that represents the person's voice biometrics can be trained in order to identify her utterances in the stored speech database.
In 2014, GoVivace Inc. started to provide speaker identification SDK for telecom companies to infer the caller's identity by matching her voice with a database containing thousands of voice recordings.
In our experiment, we can identify one person out of 250 candidates 
with a 100\% success rate given that we have collected several speeches of this person (as short as 30 seconds).
The detailed settings of this experiment will be presented in Section \ref{setup}.
Once a person's utterances is identified, the \sp will be able to mine more private information from the speech content.

Furthermore, the voice  \sp may also invade users' \textit{content privacy}. After executing \sr, 
the \sp can extract personal information from user's utterances using natural language processing techniques.
Some of the information within the utterances could be sensitive, \eg demographic attributes, personal preferences, schedules, and interpersonal relations.
Thus, the content privacy of users' speech data could be disclosed. 
Such privacy issue is always a serious concern when we use any cloud-based \sr service.

Although these two types of privacy concerns are quite different, they may lead to a worse privacy breach if combined together. For example,
with detected sensitive contents (such as phone number or home address) from the speech recognition, the speaker can be easily identified. 
even from a huge candidate pool. 
Therefore, we have to address these two types of privacy risks simultaneously.

\section{Solution Overview}
\label{sec:solution}
%


\subsection{System Architecture}
A sketch of the privacy-preserving voice input system is depicted in Fig. \ref{scenario2}. The cloud is a server dedicated to providing speech-to-text
service. It does not install apps on the mobile end. Instead, the \vs (like a keyboard app) bridges the communication between the user, the cloud, and third party apps. 
It is trusted, \eg it can release the source code to the public. 
 The \vs accesses the raw audio, perturbs it, and 
produces sanitized audio which contains much less private information about the user. After the sanitized speech is sent to the cloud and the corresponding transcript is sent back, the \vs revokes some of previous perturbations on it and restores it to the desired transcript. Finally, the transcript is copied to the input box of any other app on the device that has requested user input before.
Notice that the \vs is locally deployed in mobile devices, and it is independent of the cloud service provider. Such separation prevents the cloud from collecting private information from the user device.

\begin{figure}[t]
  \centering
  \includegraphics[width=0.48\textwidth]{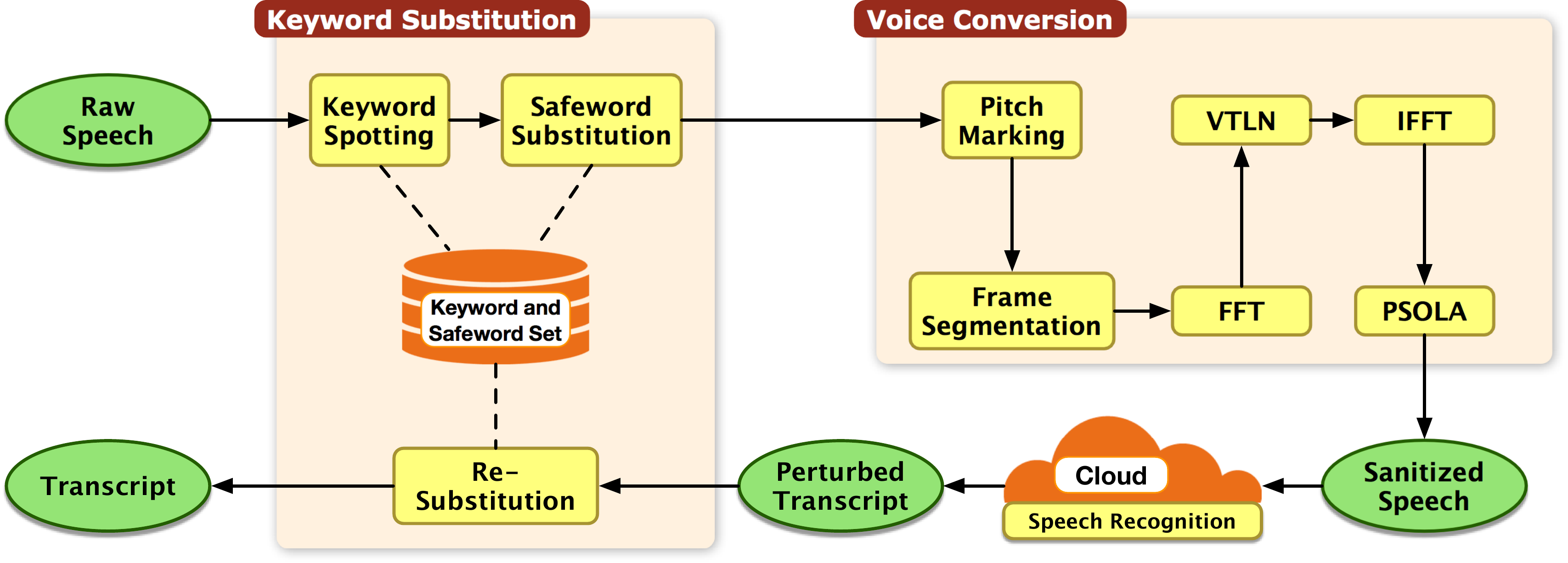} 
  \caption{\scriptsize{\textbf{The internal architecture of \VoiceMask.}}}
  \label{system_arch}
\end{figure}

The internal system architecture of the \vs is illustrated in Fig. \ref{system_arch}. Given a stream of raw audio signals, the sanitization process can be
divided into two modules: keyword substitution and voice conversion. The whole process can be explained with a simple example. 
Suppose a user is writing a text message to a friend. She reads a sentence to the mic, ``I'd love to, but I have to attend a group therapy tomorrow''. 
Suppose she has set ``group therapy'' as a sensitive keyword previously, the \vs detects its occurrence 
within the speech and then substitutes it with a selected
safeword ``meeting''. Then, the \vs converts the speech to another voice and sends the sanitized speech to the cloud. 
The cloud performs \sr and sends back the perturbed transcript ``I'd love to, but I have to attend a meeting tomorrow''.
Finally, the \vs reverses the substitution and copies the desired text to the input box of the SMS app.
In this process, the sensitive information contained in the original speech is not disclosed to the cloud. Nor is the user's voice characteristics,
which otherwise might be utilized for linkage attacks or spoofing attacks.
This is only an illustrative instance. More issues and detailed solutions will be introduced later.

\subsection{Adaptive Keyword Substitution}
Protecting content privacy is very challenging as content privacy is hard to define, let alone protecting it on unrecognized speeches.
To achieve this goal, we focus on a simplified case: sensitive keyword substitution on the speeches.
As shown in Fig~\ref{system_arch}, the process of keyword substitution contains two steps -- keyword spotting and safeword substitution.
Our approach first detects the specified sensitive keywords and then replaces them with carefully selected safewords.
Here, the sensitive keywords are predefined by the users. They could be nouns like ``hospital'' and ``addiction'', verbs like ``smoke'' and ``bribe'', adjectives
like ``alt-left'' and ``alt-right''.
The user needs to read each keyword for at least one time to train the \vs, 
and the STFT features of these utterances (referred to as \textit{keyword samples}) will be kept for keyword spotting.
The set of safewords is predefined by our system and identical to all users.
The keyword spotting and safeword substitution must be done in real-time with light computation cost.
We achieve such goals by limiting the number of keywords defined by users and a simple DTW (dynamic time warping) based method. 

\textbf{Keyword spotting} 
  identifies keywords directly from the utterances. It is usually used for keyword monitoring, speech document indexing, and automated dialogue systems \cite{chen2014small, zhang2009unsupervised}.
We design an adaptive keyword spotting mechanism based on DTW, which is well suited for real-time scenarios.
The general idea is straightforward. 
The \vs scans the input audio stream and computes the distance between the audio signal within a sliding window and the sensitive keyword samples recorded by the user previously. 
A DTW algorithm is applied here to calculate the distance of two signals, which can address the issue of dynamic speak rate and various audio lengths. 
STFT features are computed and cosine distance is used as the metric. 
If the distance between the signal of the speech segment being scanned and the signal of the keyword sample is below a certain threshold $\theta$, the corresponding segment is considered as a sensitive keyword.

However, to realize this idea, there is one more obstacle:
liaison, a very common phenomenon in English pronunciation.
The existence of liaison enlarges the distance between two identical words if they occur in different contexts. For example, the word ``apple'' can be pronounced as ``napple'' in the context of ``an apple''. Thus, directly comparing the extracted voice signals without considering liaison would lead to a high false negative rate. 
To handle this issue, we introduce an evolutionary model for representing the keywords.
For a given keyword, the stored keyword sample evolves over time. When the user speaks a sentence containing this keyword and it is correctly detected by our app (a \textit{hit}), the stored keyword sample will update itself with this new audio sample.
(The \vs would know a detected keyword is indeed a hit if the user does not revise the text recognized by it.)
Let $x$ be the feature vector of a stored keyword sample, $y$ be the feature vector of the $i$-th correctly detected keyword, 
and $\phi(y)$ represents the warped feature vector after DTW processing. To update the existing keyword sample, each element of $x$ is updated by the
following equation. 
\begin{equation}
\label{eq:refresh_key}
x'(k) = \cfrac{i}{i+1} \cdot x(k) + \cfrac{1}{i+1} \cdot \phi_y(k).
\end{equation}
This updating process mitigates the impact of liaison by laying more emphasis on the middle part.
We also allow users to point out the missed keyword and improve the corresponding recorded keyword in the same way.

\textbf{Safeword substitution:} 
After spotting the keyword, we replace it with a word chosen at random from the safeword set.
\Vs will keep the log of such replacement, which enables  re-substituting the safeword back after receiving the perturbed transcript from the cloud.
This randomized substitution also avoids the privacy leakage from choosing safeword. We will present more details in Section \ref{content}.

\subsection{VTLN-Based Voice Conversion}
\label{vtln}

\begin{figure}[t]
  \centering
 \subfigure[Bilinear functions]{\includegraphics[width=0.46\linewidth]{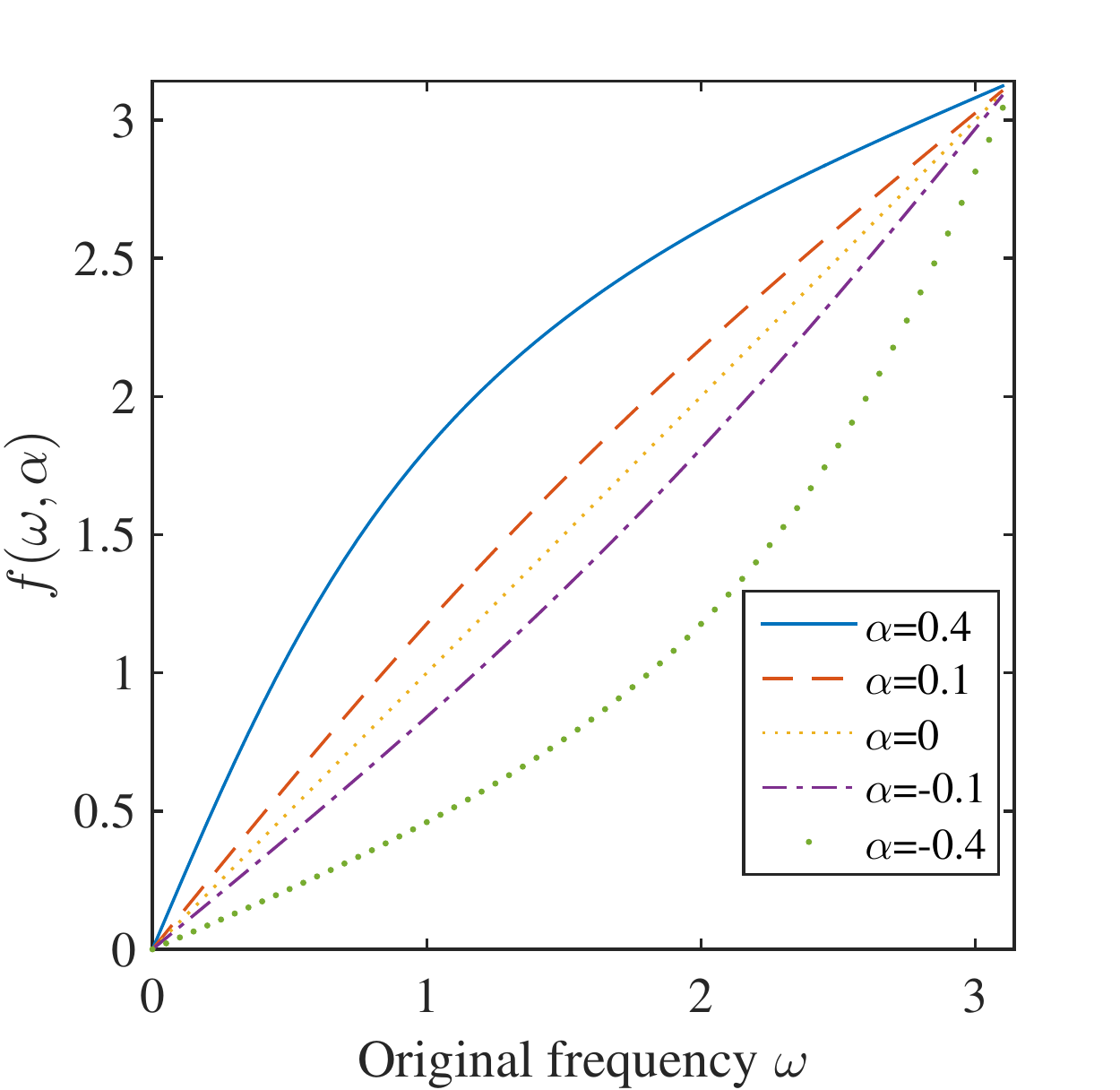}}
 \subfigure[Example of effect]{\includegraphics[width=0.50\linewidth]{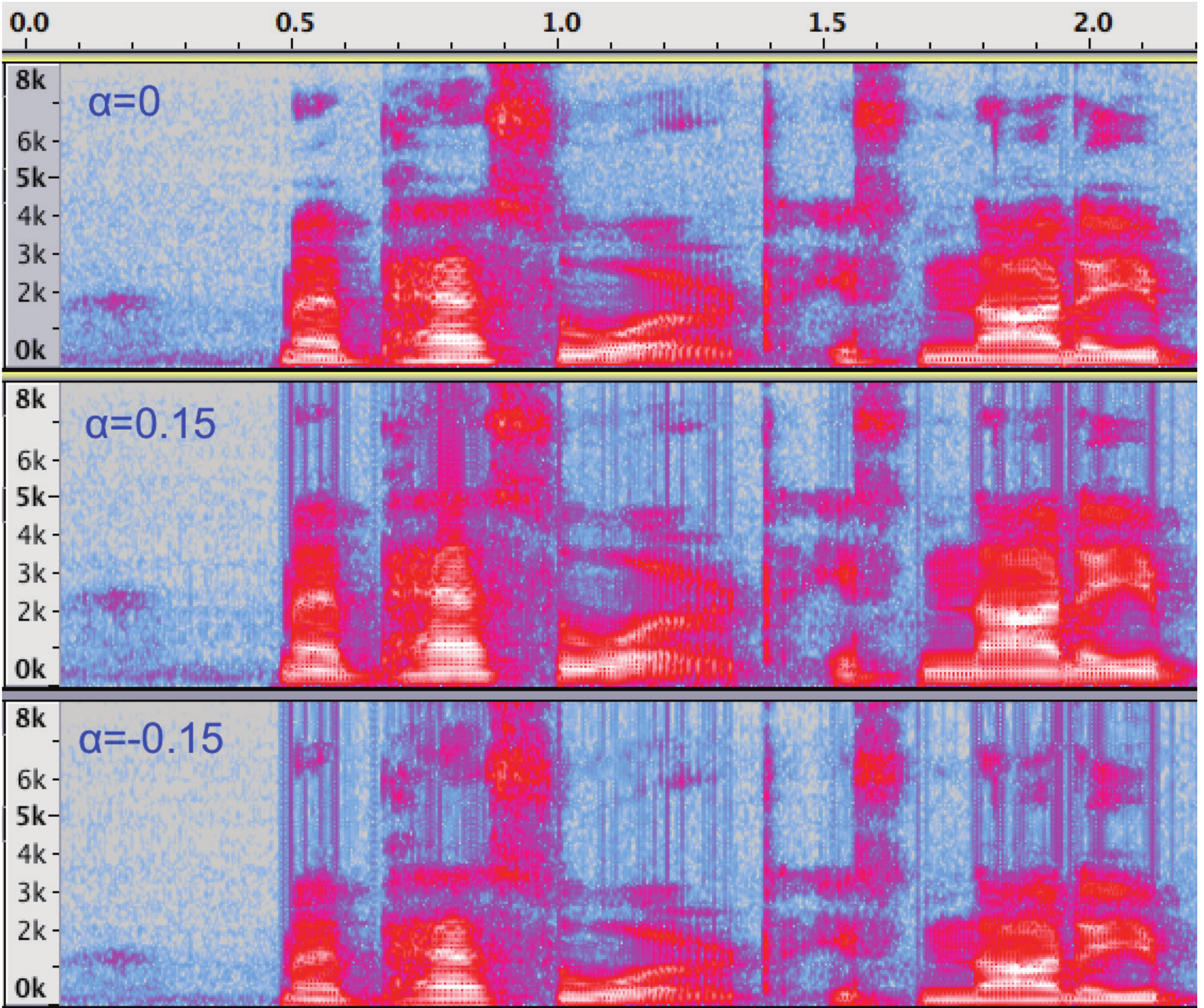}}
 \vspace{-0.1in}
  \caption{\scriptsize{\textbf{Bilinear functions:} A bilinear function is monotone increasing in the domain $[0,\pi]$ and the range $[0,\pi]$. When $\alpha<0$, the low frequency part of the frequency axis is compressed and the high frequency part is stretched. When $\alpha>0$, the low frequency part is stretched while the high frequency part is compressed. Fig. (b) gives an example of the spectrogram of a speech before and after its frequencies are warped.}}
  \label{bilinear}
\end{figure}

To protect identity privacy for users of voice input service, we design a mechanism on top of the voice conversion technology.
Voice conversion algorithm modifies a source speaker's voice so that it sounds like another target speaker without changing the language contents. One of the most popular voice conversion paradigms is frequency warping  \cite{sundermann2003vtln},
which is based on the well-studied technique of vocal tract length normalization (VTLN) \cite{eide1996parametric,cohen1995vocal}.
Given a source utterance, VTLN-based voice conversion processes it in 6 steps (Fig. \ref{system_arch}).
Pitch marking and frame segmentation aim to split the speech signal into frames that match the pseudo-periodicity of voiced
sounds as determined by the fundamental frequency of the voice, so as to make the output synthetic voice have the best audio quality.
The key step of voice conversion is VTLN, which modifies the spectrum of each frame using frequency warping,
that is, stretching or compressing the spectrum with respect to the frequency axis according to a warping function.
One of the most commonly used warping function is the bilinear function \cite{sundermann2003vtln,acero1991robust}. 
The formula of this function is:
\begin{equation}
\label{equ:bilinear}
f(\omega,\alpha)=\vert -i\ln\frac{z-\alpha}{1-\alpha z} \vert,
\end{equation}
where $\omega\in[0,\pi]$ is the normalized frequency, $\alpha\in(-1,1)$ is a warping factor used to tune the strength of \vc, 
$i$ is the imaginary unit and $z=e^{i\omega}$.
Several examples of the bilinear function are plotted in Fig. \ref{bilinear}.
Given a frequency-domain data frame, every frequency $\omega$ is changed to a new frequency $f(\omega,\alpha)$ by this formula.
Then all the frequency-warped frames are reverted to the time domain by IFFT.  Finally, 
 all the frames are concatenated to generate the output speech, and a technique 
 PSOLA is utilized to improve the speech quality \cite{valbret1992voice}. 
Therefore, an utterance is produced with the same language content but a different voice.
This is the basic idea of protecting the identity privacy, yet it is not that easy.  There are several issues to consider, including how to
minimize the influence on the audio quality of the output speech and how to select $\alpha$ to achieve a strong privacy guarantee.
We will discuss the details in Section \ref{identity}.

\section{Identity \& Content Privacy} 
\label{sec:content}
%
\subsection{Protecting User Identity Privacy}
\label{identity}
When we disguise the speaker's identity by randomly modifying her/his voice to another voice, it is required that
the speech content can still be accurately recognized when it is uploaded to the cloud so that user experience is not degraded unacceptably.
As aforementioned, 
the parameter  $\alpha$ in the bilinear function tunes the extent of distortion of the output voice.
Setting $\alpha<0$ would produce a deeper (more like low-pitched) output voice;
setting $\alpha>0$ would produce a sharper (more like high-pitched) output voice.
The greater $|\alpha|$ is, to a higher extent the output voice is distorted.
The output voice is not distorted at all when $\alpha$ is 0. (See Fig. \ref{bilinear} for the reason.)
We want to select the best $\alpha$ that can bring a considerable drop in the
\spk accuracy whereas the decrease of the \sr accuracy is minimized.
However, finding the best value of $\alpha$ is not enough, because
if we always use the same $\alpha$ to transform a user's voice to the same fake voice, 
the cloud may build a model on this fake voice and use it to recognize this user afterward.
In addition, the cloud may find the value of $\alpha$ by decompiling the apk of the \vs, and then reverse the voice conversion to recover the original voice.
\textit{Reversing voice conversion} attack is possible because the warping functions are invertible. Take bilinear functions as an example, 
$f(w,-\alpha)$ is the inverse function of $f(w,\alpha)$. 
Therefore, we need to find a \textit{proper range} of $\alpha$ and randomly choose $\alpha$ from this range every time.
Now the question is, what is the proper range and how randomly should we choose $\alpha$ from it to ensure reversing voice conversion is impossible?

\textbf{Randomized warping factor selection:}
Suppose the proper range of the warping factor $\alpha$ is $A$. In fact, we found $A=[-0.10,-0.08]\cup[0,08,0.10]$ 
 experimentally (we will present details in Section \ref{vc_effect}).
Every time the user speaks a sentence, the \vs randomly selects $\alpha\in A$,
uses Eq.~\eqref{equ:bilinear} to warp the frequency of the speech and then sends it to the cloud. 
The cloud receives speeches in different voices even when they are from the 
same user, so it is almost impossible to directly use the previously trained voice models to recognize the user's identity.

However, there is still a possibility that the cloud can reduce (``partially reverse'') the voice conversion 
to a weaker level so that it can achieve a higher \spk accuracy. 
The bilinear function $f(\omega, \alpha)$ we adopt has a property: 
\begin{equation}
\label{equ:reduce+}
f(f(\omega, \alpha_1), \alpha_2)=f(\omega, \alpha_1+\alpha_2).
\end{equation}
That said, applying \vc twice to a speech with $\alpha_1$ and $\alpha_2$ successively 
yields exactly the same output as applying the voice conversion once with $\alpha=\alpha_1+\alpha_2$.
Suppose the cloud has received many sanitized speeches from users whose voices have been converted with $\alpha\in [0.10,0.12]$.
It can apply a second \vc to these speeches with $\alpha_2=-0.11$.
Now the produced speeches are actually the output speeches of \vc with $\alpha\in[-0.01,0.01]$, 
which has much weaker distortion strength than the originally sanitized speeches.
Consequently, the cloud can achieve better accuracy than we originally expected when performing \spk on these speeches.
We refer to this process as \textit{reducing \vc} attack, and we say a function is \textit{reducible} if it has the property in 
Eq.~\eqref{equ:reduce+}. To our knowledge, this issue has not been studied in previous research papers.
Besides the bilinear function, other common warping functions like the symmetric function \cite{uebel1999investigation}, 
the power function \cite{eide1996parametric} and the quadratic function \cite{pitz2005vocal} are also reducible.
Therefore, a more complex mechanism must be designed.

\textbf{Secure compound warping functions:}
To design a warping function that is resistant to the reducing \vc attack, our technique is to compound two different warping functions.
Here we introduce another commonly used warping function that is easy to tune, the quadratic function \cite{pitz2005vocal}:
$g(\omega,\beta)=\omega+\beta (\dfrac{\omega}{\pi}-(\dfrac{\omega}{\pi})^2)$,
where $\omega\in[0,\pi]$ is the normalized frequency and $\beta>0$  is a warping factor.
Similar to $\alpha$ in the bilinear function, $\beta$ determines the distortion strength of this function.
The output voice turns deeper when $\beta<0$ and sharper when $\beta>0$.
Luckily, the compound function of $f$ and $g$ (denoted as $h$) is not reducible.
The formula of $h$ is: 
$h(\omega, \alpha, \beta)=g(f(\omega,\alpha),\beta)$.
Since the two independent parameters $\alpha,\beta$ are used in combination, they have a much bigger proper range (will be studied in Section \ref{para_select}).
Now if every time the \vs perturbs a speech from the user, it randomly picks a pair of values for $\alpha,\beta$ from this range as the warping factors.
This mechanism ensures the cloud is unable to reverse or reduce the \vc.

However, it is challenging to learn the combined impact of $\alpha,\beta$ on the distortion level of the output voice.
To estimate it, we need to first quantify the \textit{distortion strength} of the warping function $h$.
An intuition is that the closer $h$ is to the identity function, the less distortion $h$ brings to the output voice.
Take Fig. \ref{bilinear}(a) as an example, the closer the bilinear function is to the identity function,
the closer $\alpha$ is to 0, and the less distortion it produces on the output voice.
This way, we can measure the distortion strength of $h$ by the area between the curves of itself and the identity function.
\begin{definition}[Distortion strength]
\label{def:dist}
The distortion strength (denoted as $dist$) of a warping function $f(\omega,\mathbf{a})$ is
defined as the area between the curves of itself and the identity function, \ie
$dist_f(\mathbf{a})= \int_0^{\pi} \mid f(\omega, \mathbf{a})-\omega\mid$,
where $\mathbf{a}$ represents the warping factors.
\end{definition}
We will discuss how to find the proper range using this definition in Section \ref{para_select}.

\subsection{Protecting Speech Content Privacy}
\label{content}
Content privacy is achieved by substituting sensitive keywords with safewords, so the first task is to construct a safeword set.

\textbf{Differentially private safeword set construction:}
We don't want to replace a keyword in a user's speech with a more sensitive word because this would harm her personal reputation,
for instance, when the keyword ``doctor'' in  ``I need to see my doctor'' is 
changed to ``probation officer''. Thus, we need to avoid putting common sensitive words in the safeword set. 
But how does the \vs know which words are usually sensitive?
We introduce a new central server that connects all the \vs apps and collects the keywords defined by users for statistics.
The server ranks these keywords by frequency and then knows which words are commonly viewed as sensitive.
Nevertheless, we cannot fully trust the server. Directly sending each user's sensitive keywords to the server would break their privacy.
Inspired by \cite{erlingsson2014rappor}, we design a privacy-preserving randomized keyword aggregation algorithm 
(PRAKA) that achieves differential privacy
based on the randomized response mechanism \cite{dwork2014algorithmic}.
Given a vocabulary $\mathcal{V}$ that contains all the possible keywords and a parameter $p$ set by the user to tune the privacy level, the \vs executes the algorithm in the following steps.
For each word $w\in\mathcal{V}$: (1) Create a 2-bit array $B=[b_1,b_2]$ and initialize it to `00'. (2) If the user sets $w$ as sensitive, then set $b_1$ to 1; otherwise set $b_2$ to 1. (3) For each bit $i$ of $B$, set it to 1 with probability $\frac{p}{2}$, set it to 0 with probability $\frac{p}{2}$, or do not modify it with probability $1-p$. Let the new array we get be $B'$. (4) Report $\langle w,B'\rangle$ to the central server.
We have the following theorem (see proof in Appendix \ref{app1}).
\begin{theorem}
PRAKA achieves $2\ln \frac{2-p}{p}$-differential privacy.
\end{theorem}

Since the user reports to the server \textit{only once} for each $w$, the privacy level would not degrade over time.
When the server collects enough reports from users, it estimates the frequency of each word set as sensitive by users. 
For each $w$, let $N$ be the total number of reports, let $n$ be the number of reports where bit 1 was set, the estimated number of users who really set 
$b_1=1$ is calculated by $\hat{n}=(n-\frac{p}{2}N)/(1-p)$. The error bound is given in Appendix \ref{app2}.

\textbf{Randomized safeword selection:}
Given a safeword set, we split it into several subsets such that words within a subset are more similar to each other than words in another subset.
For instance, a simple categorization is singular nouns, plural nouns, transitive verbs, intransitive verbs, adjectives, and adverbs. 
Each subset is a bucket of safewords.
The categorization is pre-defined in the app. 
When the \vs spots an occurrence of a keyword, 
 it randomly picks a safeword from the corresponding bucket to perform safeword substitution.
\section{Evaluation}
\label{sec:evalution}\label{sec:emulation}

\subsection{Emulation Setup}
\label{setup}

We run the emulation on three \textbf{datasets}.
\textit{PDA}: PDA is a speech database from CMU Sphinx Group \cite{pda},
which contains 16 speakers each speaking over 50 short sentences.
The number of female and male speakers are well balanced.
They are all native English speakers, and 12 of them speak American English.
%
 \textit{LibriSpeech}: LibriSpeech corpus is from OpenSLR \cite{panayotov2015librispeech},
containing 100 hours speech, clearly read by 251 native speakers of  American English without background noise.
 \textit{Volunteers}: This dataset was collected by ourselves from 14 college students (11 males and 3 females).  
Thirteen of them are not native speakers and they have various accents from 9 countries. 
Their statistics are shown in Tab. \ref{tab:real-data}.

\begin{table}\scriptsize
\centering
 \caption{\scriptsize{\textbf{Statistics of the datasets in our evaluation.} }}
 \vspace{-0.1in}
\label{tab:real-data}
 \begin{tabular}{ c | c | c | c | c }
    \hline
    Dataset &  \#Speakers &  \#Speeches & Hours & English accents  \\ \hline \hline
    PDA & 16 & 836 & $1.8$h & Mostly native \\  \hline
    LibriSpeech & 251 & 27.7k & $100$h & All native  \\ \hline
    Volunteers & 14 & 240 & $0.7$h & Various accents  \\ \hline
 \end{tabular}
\vspace{-0.1in}
\end{table}

We utilize the Speaker Recognition API and the Bing Speech API from Microsoft Cognitive Services, which
provides the state-of-the-art algorithms for spoken language processing~\cite{micWER, xiong2016achieving}. 
The \textbf{steps of evaluation} go as follows.
First, we create a speaker recognition system
and use the training set to train a voice model for every speaker in the three datasets.
Second, we process the utterances in the test set using our \vs.
We perform voice conversion differently for male and female speakers. Specifically,
we deepen female voices (by setting $\alpha<0$) and sharpen male voices (by setting $\alpha>0$) to make their voices closer in pitch
so as to increase the difficulty for the adversary in distinguishing different speakers (validated by experiment).
Third, we use the trained voice models to identify the speakers of the sanitized utterances and evaluate the accuracy of speaker identification.
Finally, we perform speech recognition on the  sanitized utterances and evaluate the accuracy.

\textbf{Metrics:}
We measure the performance of \sr using word accuracy, which is calculated by one minus word error rate (WER).
The accuracy of \spk is defined as the success rate, that is,  the fraction of correctly identified speakers.
By default, we measure the accuracy of identifying a speaker from a set of 50 candidates, but we also study the cases where there are much more candidates (Fig. \ref{spk_acc_ncandidates}).
We measure the performance of \ks using ROC curve, which illustrates the performance of a binary classifier system as its discrimination threshold is varied.
The computation overhead of the \vs
 is measured by \textit{realtime coefficient}, the ratio between the CPU time for processing the audio and the duration of the audio.

\subsection{Emulation of Voice Conversion}
\subsubsection{Effectiveness}
\label{vc_effect}
\begin{figure}[t]
  \centering
\subfigure[PDA dataset]{\label{acc_spk_sr}\includegraphics[width=0.48\linewidth]{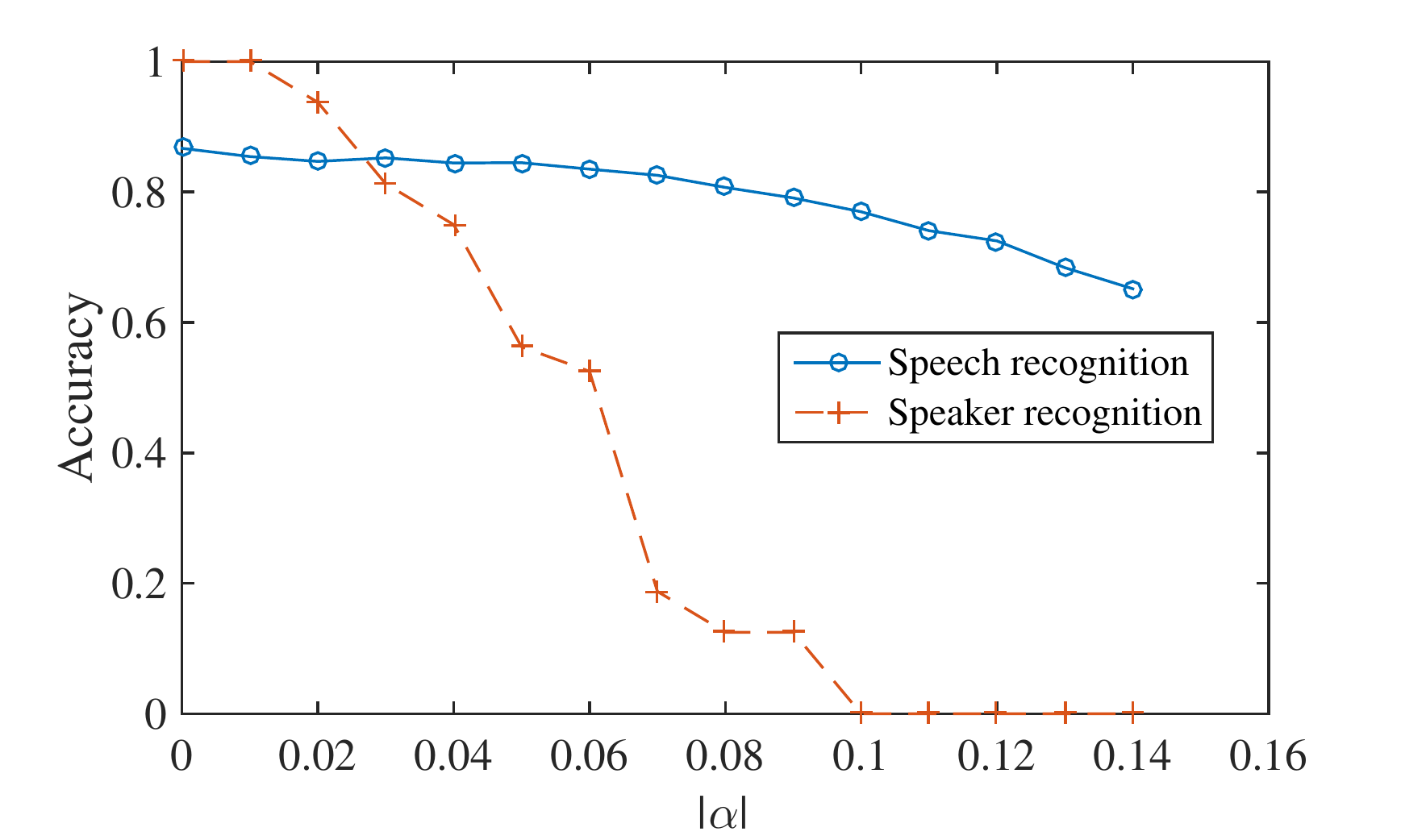}} \ 
\subfigure[LibriSpeech dataset]{\label{acc_spk_sr2}\includegraphics[width=0.48\linewidth]{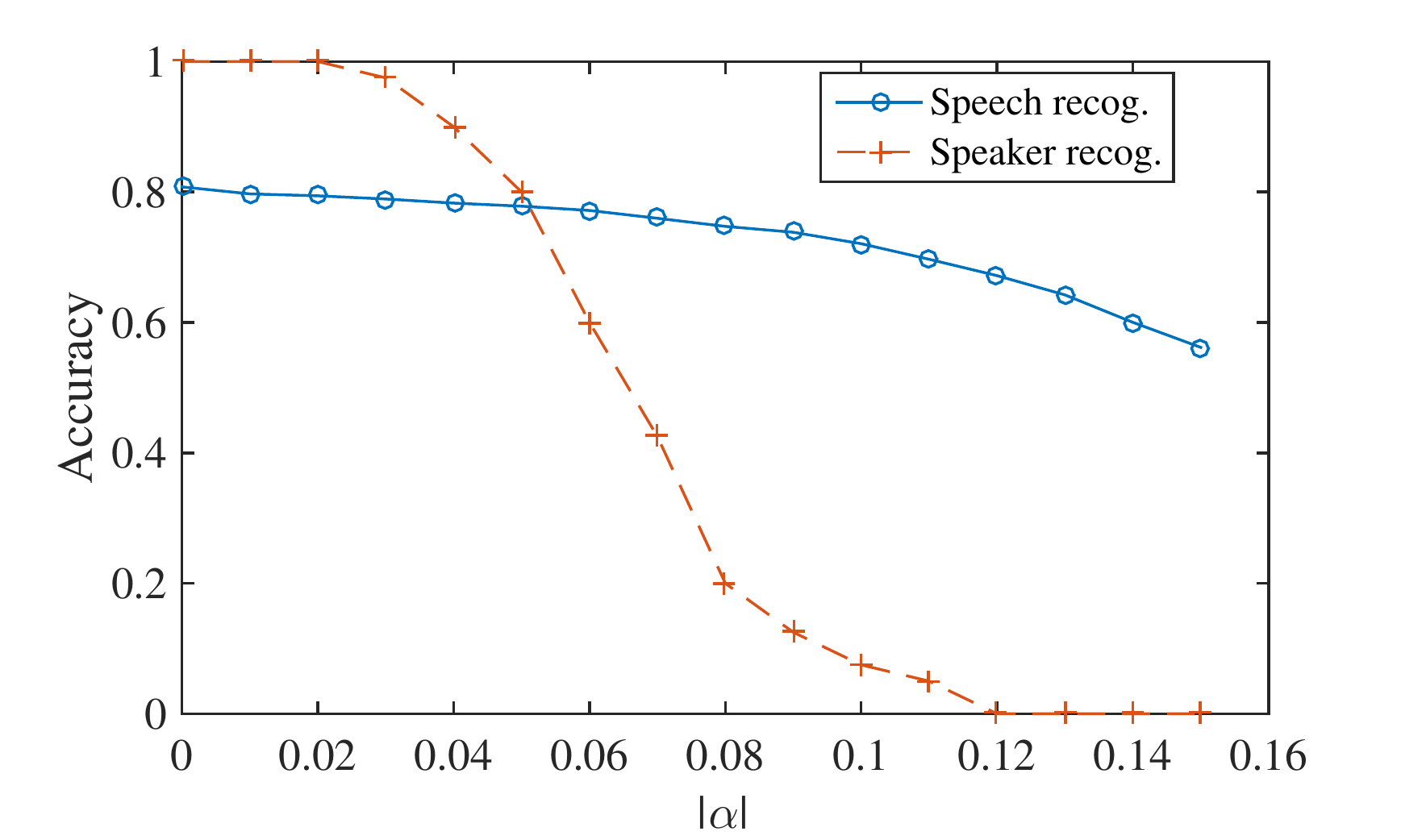}} 
\vspace{-0.1in}
  \caption{\scriptsize{\textbf{The impact of voice conversion on the accuracy of speaker/speech recognition.}  When $|\alpha|\in [0.02,0.10]$, the accuracy of \spk goes down significantly.}} 
  \label{sr_spk_acc}
\end{figure}

Both \sr and \spk have accuracy degradation when the utterances are processes by our \vs,
but the extent of degradation with respect to a specific $|\alpha|$ is different for them.
We can observe that in Fig. \ref{sr_spk_acc}.
Speakers can be correctly identified with a 100\% chance when $\alpha=0$, \ie on original utterances.
Speech recognition on the original utterances achieves an accuracy $>80\%$. 
When voice conversion is applied with $|\alpha|\in [0.02,0.10]$,
the accuracy of speaker recognition degrades sharply with growing $|\alpha|$
while that of \sr is barely influenced.
The huge gap provides us an opportunity to find a value range of $|\alpha|$ to overcome our key challenge, that is,
to suppress the speaker recognition possibility while
preserving the speech recognition utility.
According to this figure, the proper range of $|\alpha|$ can be set to $[0.08,0.10]$.
That said, if we use only the bilinear function in the \vc for LibriSpeech and choose $|\alpha|$ from $[0.08,0.10]$,
the accuracy of \spk on the output speeches would be restricted within $0.20$ but the \sr accuracy is still maintained at $0.72\sim 0.75$.
The results for PDA are even better: the \spk accuracy is restricted to be smaller than $0.13$ while the \sr accuracy is in $0.77\sim 0.81$.

\begin{figure}[t]
  \centering
  \includegraphics[width=0.6\linewidth]{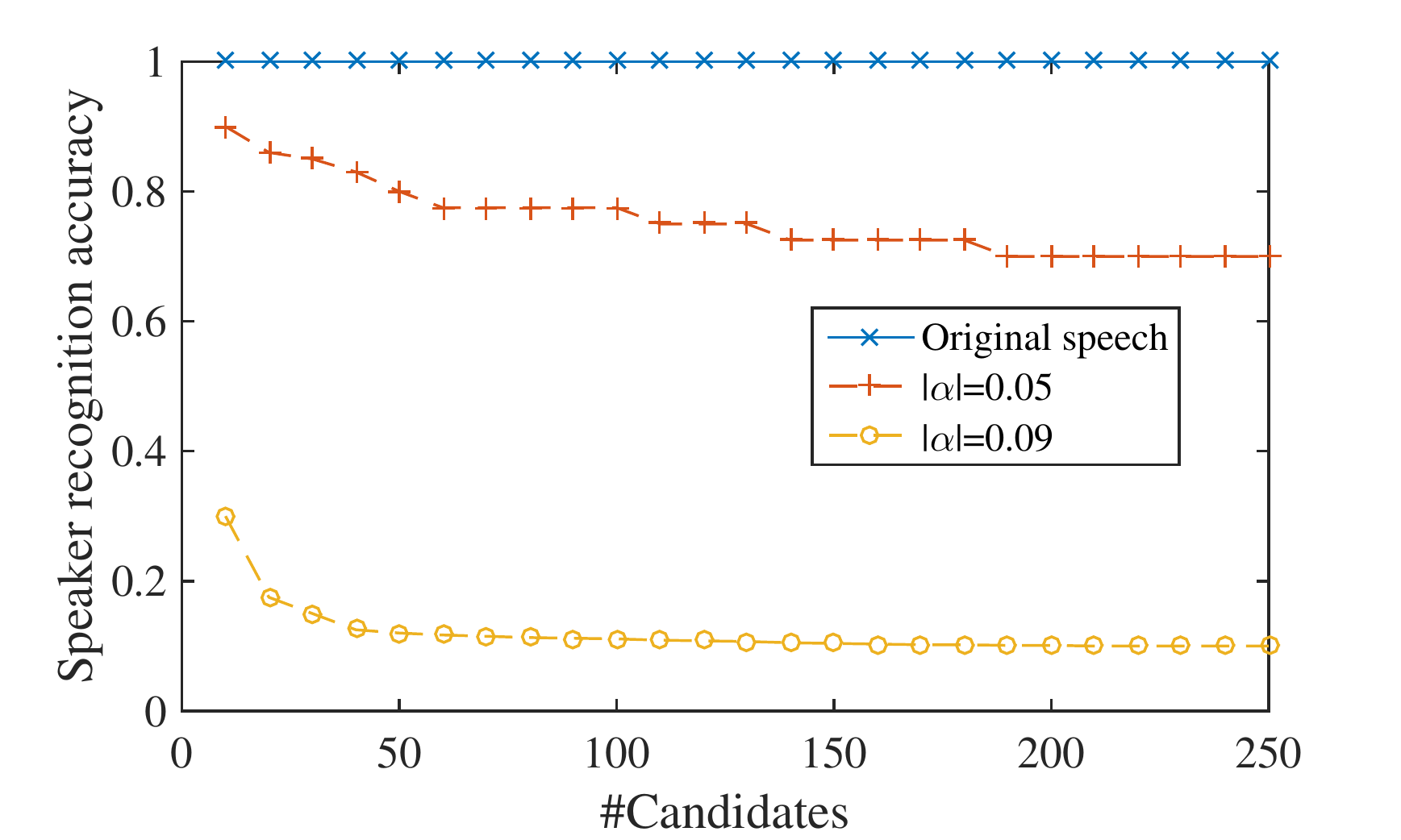}
\vspace{-0.1in}
  \caption{\scriptsize{\textbf{Speaker recognition accuracy vs $|\alpha|$ and the number of candidates (on LibriSpeech).} The accuracy of \spk decreases when $|\alpha|$ increases or when there are more candidates.}} 
  \label{spk_acc_ncandidates}
\end{figure}

The accuracy of \spk is also partially up to the number of candidates from which the cloud identifies the target speaker,
as shown in Fig. \ref{spk_acc_ncandidates}.
However, if we do not apply voice conversion, the accuracy decreases extremely slow such that it is still 1 when we identify the speaker from 250 candidates.
(We could not try more candidates because of the limit of dataset we obtained.)
This figure shows that our method can reduce the speaker recognition accuracy by 90\%.

\vspace{-0.1in}
\subsubsection{Computational Cost}
We record the CPU time of voice conversion on every utterance.
The time of pitch marking and other 5 steps, and the total CPU time are all proportional to the duration of the utterance.
Pitch marking is the most time-consuming step.
The realtime coefficient of voice conversion is 0.42 on average (Tab. \ref{tab:realtime_coefficient}).
We also find that the warping parameter $\alpha$ has little influence on the computational cost.

\subsubsection{Emulation of Robust Voice Conversion}
\label{para_select}

\begin{figure}[b]
\vspace{-0.3in}
  \centering
  \includegraphics[width=0.48\textwidth]{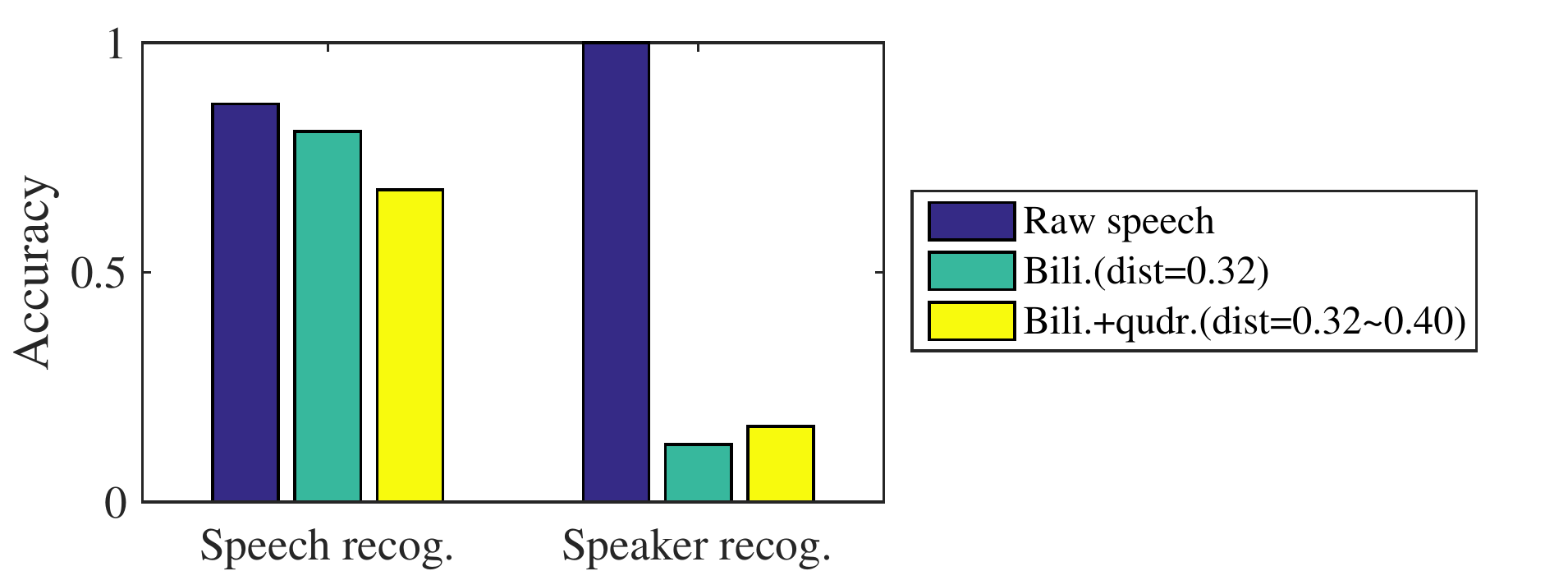}
  \vspace{-0.1in}
  \caption{\scriptsize \textbf{Comparison of different voice conversion techniques.} We compound the bilinear and quadratic functions to improve the robustness, though the performance is not as good as using only bilinear function.}
  \label{acc_compound}
\end{figure}
\begin{figure}[t]
  \centering
\subfigure[When $dist_h\leq 0.40$]{\label{range} \includegraphics[width=0.48\linewidth]{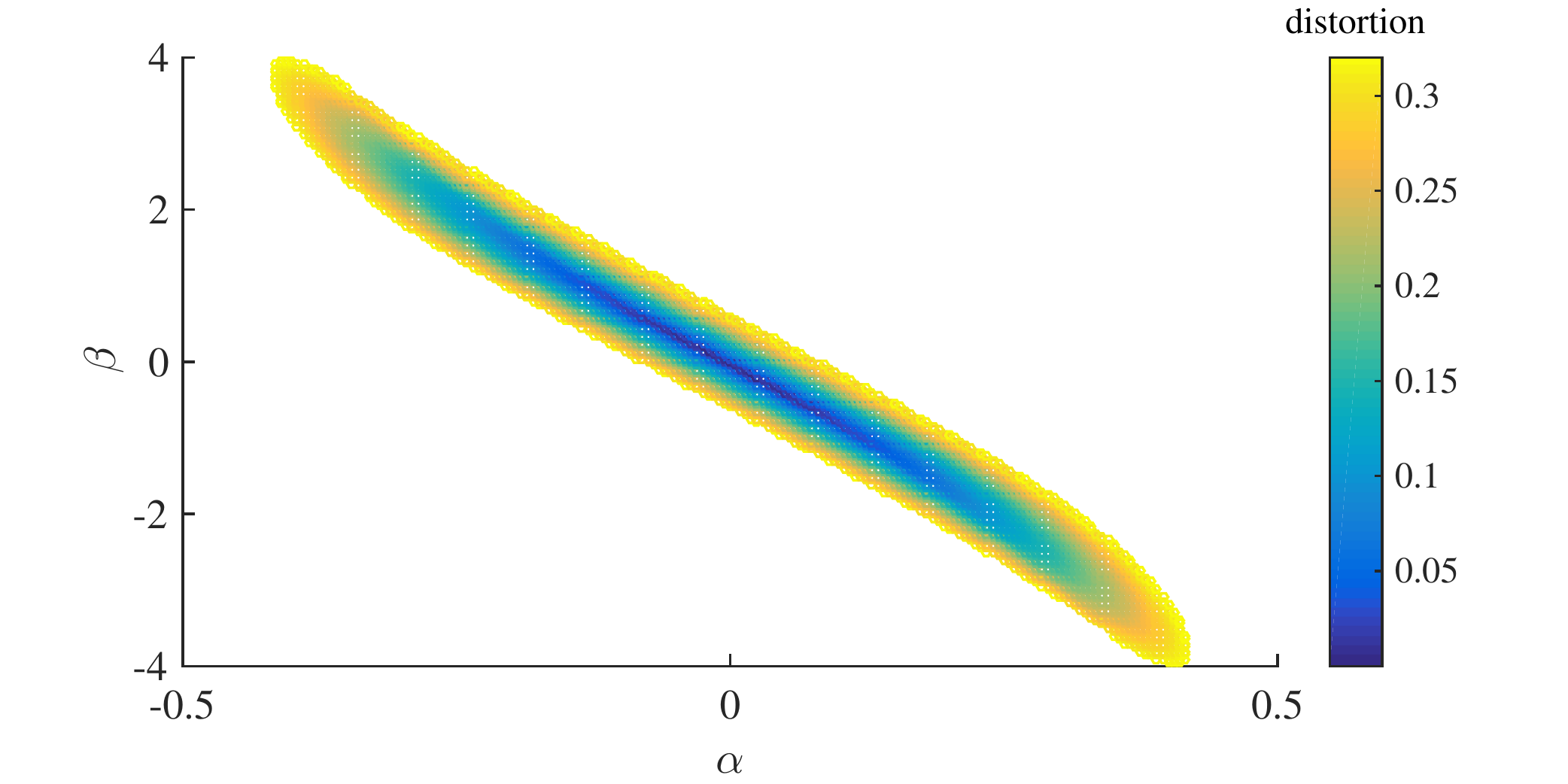}}
\subfigure[When $0.32\leq dist_h\leq 0.40$]{\label{properRange} \includegraphics[width=0.48\linewidth]{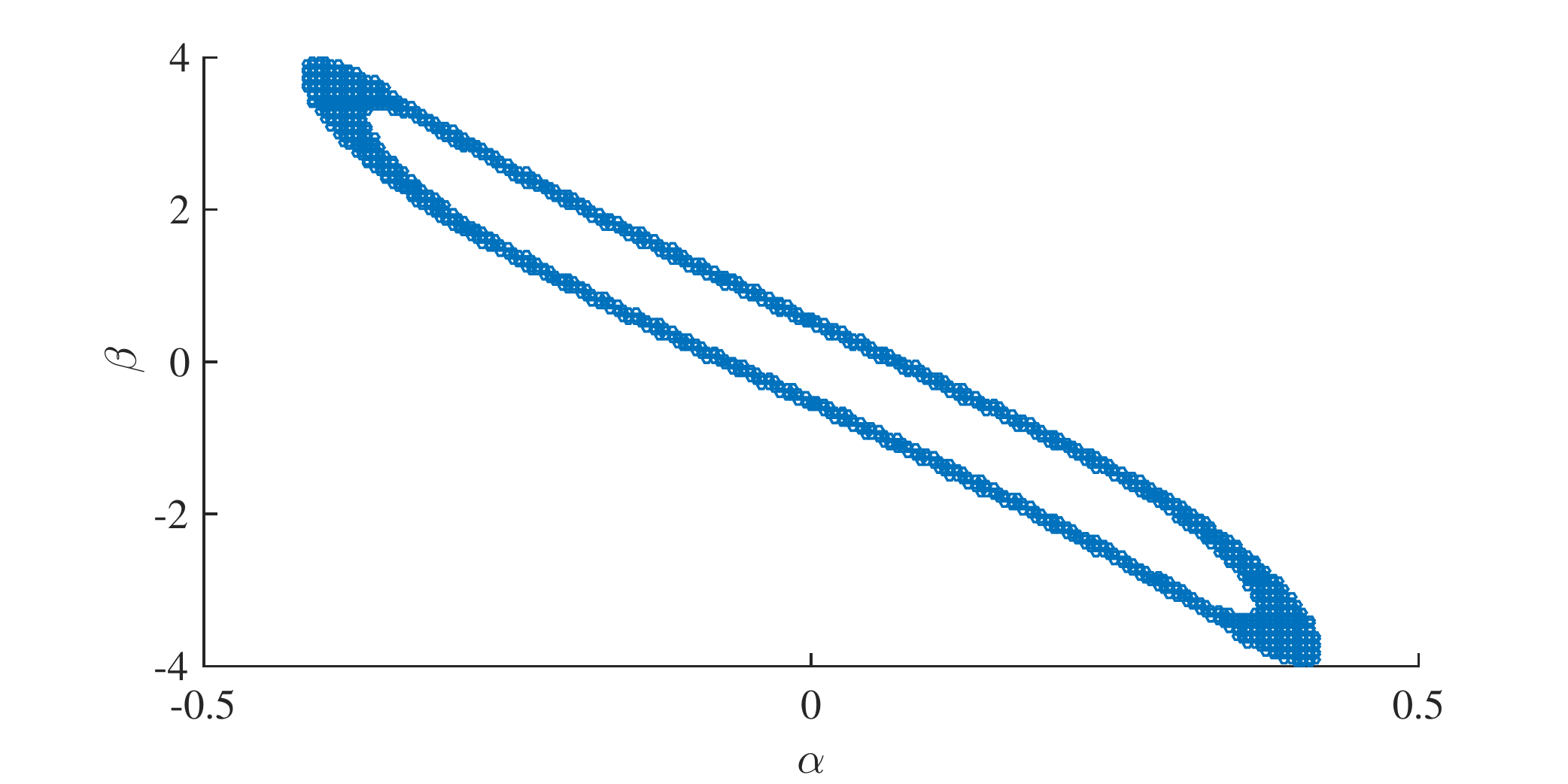}}
  \caption{\scriptsize{\textbf{The impact of $\alpha,\beta$ on voice distortion and their proper range.}}}
  \label{distortion_range}
\end{figure}
We have presented the idea of adopting a compound function
$h(\omega, \alpha, \beta)$ in Section \ref{identity}, so now we try to find the proper range of $\alpha,\beta$ for this function.
By Definition \ref{def:dist}, the distortion strength of our compound function $h(\omega, \alpha, \beta)$ can be estimated. 
Now the question is, what is the boundary of a \textit{proper} distortion strength? 
As mentioned in Section \ref{vc_effect}, the proper range of $|\alpha|$ in the bilinear function $f$ is $[0.08,0.10]$.
The corresponding distortions are $dist_f(0.08)=0.32$, $dist_f(0.10)$ $=0.40$.
So we adopt these two values as the boundary of a proper distortion: $dist\in [0.32,0.40]$.
Fig. \ref{properRange} shows the proper range of $\alpha, \beta$ for $h$ such that $0.32\leq dist_h(\alpha, \beta)\leq 0.40$.
By randomly selecting $\alpha,\beta$ from this range every time we convert the voice of a speech, we can achieve our goal of
impeding \spk while preserving \sr, plus the goal of preventing the reducing \vc attack.
An experiment on PDA demonstrates the efficacy of this technique, as presented in Fig. \ref{acc_compound}.
The test on the unsanitized speech (dark blue bars) and weakly sanitized speech (green bars, bilinear function with $\alpha=0.08$)
is also given as a contrast.
After the speeches are perturbed by \vc with the compound warping function, the speech/speaker recognition accuracy decreases to 0.68 and 0.16 respectively,
as indicated by the yellow bars. We can improve speech recognition accuracy by lowering the privacy level.

\subsection{Emulation of Keyword Substitution}
\begin{figure}[b]
\vspace{-0.3in}
  \centering
  \includegraphics[width=0.27\textwidth]{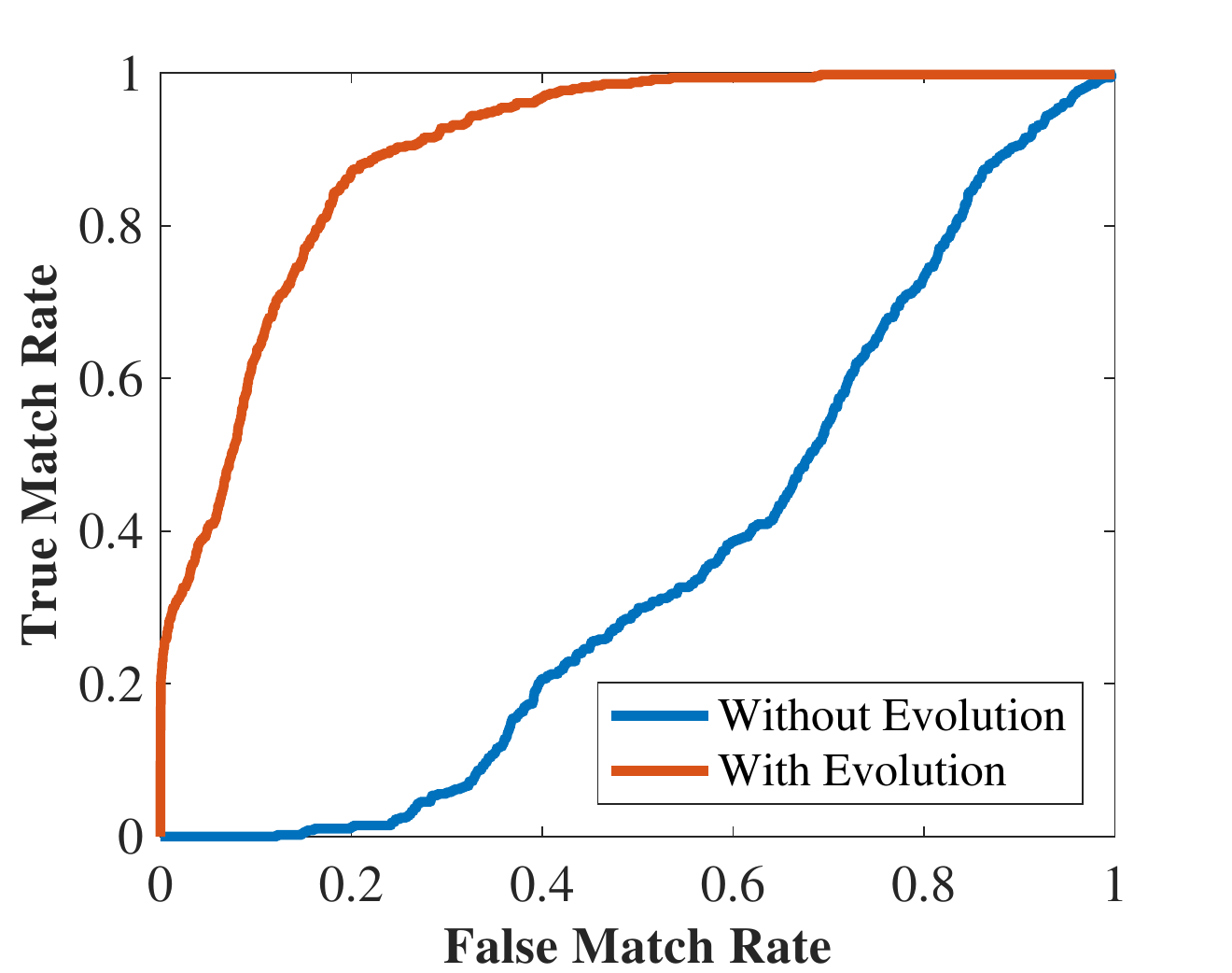}
  \vspace{-0.1in}
  \caption{\scriptsize{\textbf{ROC curve of (evolution-based) keyword spotting.} }} 
  \label{fig:combined_ks}
\end{figure}
\begin{table*}[t]\scriptsize 
\centering
 \begin{tabular}{ c | c | c | c | c | c | c | c }
    \hline
    \multirow{2}*{} & \multicolumn{3}{c |}{Voice conversion} & \multicolumn{3}{c |}{Keyword substitution} &  \multirow{2}*{Total} \\\cline{2-7}
    \multirow{2}*{} & Pitch marking & Other steps & Subtotal & Keyword spotting & Substitution & Subtotal & \multirow{2}*{}   \\ \hline \hline
    Real time coefficient & 0.35 & 0.07 & 0.42 & 0.56  & 0.04 & 0.60 & 1.02 \\ \hline
    Standard deviation & 0.05 & 0.01 & 0.05 & 0.07  & 0.01 & 0.07 & 0.07  \\ \hline
 \end{tabular}
 \caption{\scriptsize{\textbf{Real time coefficient of voice sanitization}}}
\label{tab:realtime_coefficient}
\end{table*}

We evaluate the performance of our keyword substitution method on LibriSpeech. 
For every user, we constructed the keyword set by randomly choosing words from the transcripts and then manually labeled the occurrences of these words among all the speeches of this user. 
We apply our evolution based method to detect the keywords, and then compare the results with the labels.
If the difference between detected start/end time and labeled time is smaller than 0.05s, this detection would be considered as correct.

As illustrated in Fig~\ref{fig:combined_ks}, without updating the keyword samples, the detection accuracy approximates to random guess.
Our evolution based algorithm helps to improve such detecting accuracy to 90\%, as indicated by the red line.
To measure the computational cost, we assume there is one keyword per utterance. The most common case is zero or one keyword because most words in a sentence are insensitive at all. The realtime coefficient of substituting one keyword is 0.60, as shown in Tab. \ref{tab:realtime_coefficient}.
The overhead is proportional to the number of keywords. With the development of \ks, we can achieve better efficiency in the future.

\begin{figure}[b]
\vspace{-0.3in}
  \centering
  \includegraphics[width=0.43\textwidth]{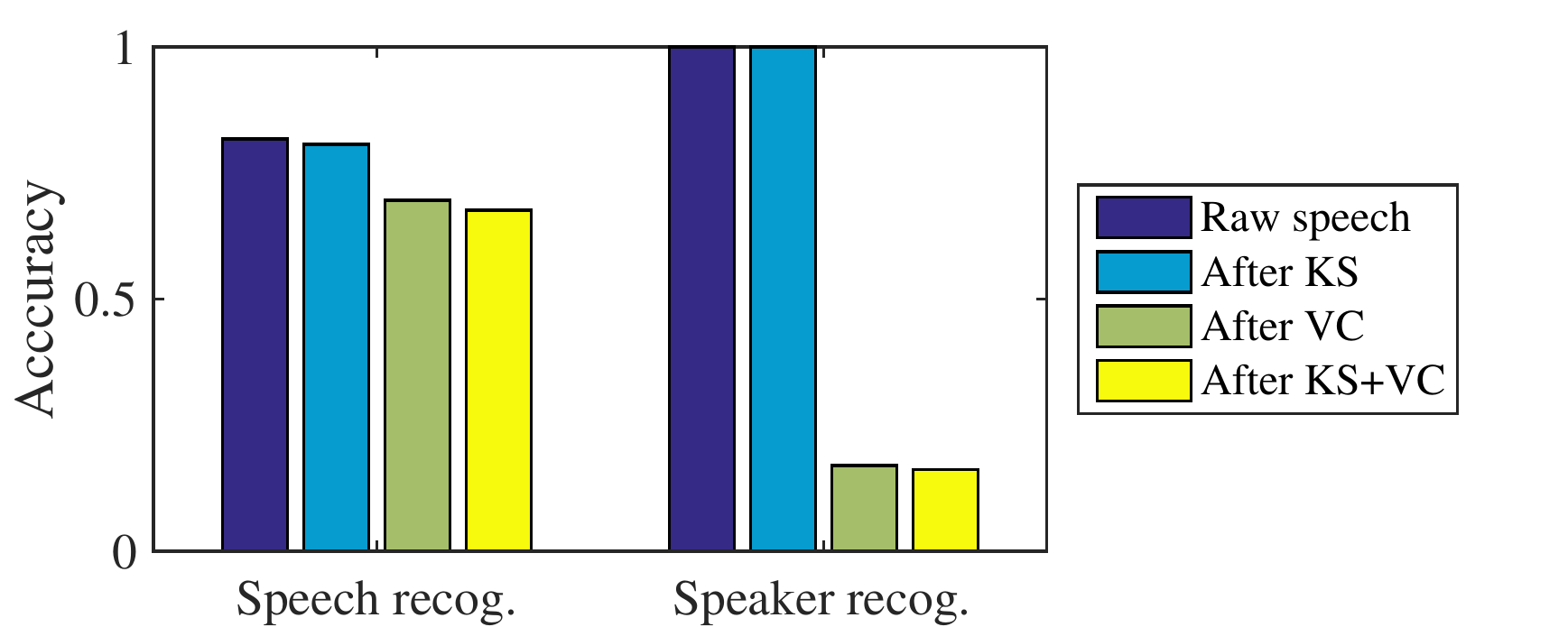}
  \vspace{-0.1in}
  \caption{\scriptsize{\textbf{The impact of keyword substitution (KS) and voice conversion (VC) on the accuracy.}}}
  \label{acc_ks_vc}
\end{figure}

\textbf{Put it all together:} We combine the two sanitization phases (keyword substitution and \vc)
altogether and run the same experiment. The impact of speech sanitization on the
accuracy of \sr and \spk is presented in Fig. \ref{acc_ks_vc}. Keyword substitution has a marginal impact on the performance of speech/speaker recognition.
It is shown that we can reduce the risk of a speaker from being identified (given 20 candidates) from 100\% to 16\%
at the cost of decreasing the \sr accuracy by 14.2\%. The computation overhead of speech sanitization is given in Tab. \ref{tab:realtime_coefficient}.
The total real time coefficient is 1.02.

\section{Discussion}
\label{sec:discussion}
\jnl{
\textbf{Other keyword spotting methods:}
\Ks can be implemented with phonetic search, acoustic methods, LVCSR-based methods~\cite{moyal2013phonetic}.
Acoustic \ks only differentiates keywords and non-keywords, 
so it is much more efficient than performing \sr on the whole speech.
Acoustic keyword spotting can be implemented with two methods, HMM/GMM
and TRAP/NN 
\cite{szoke2005comparison}.
Ideally we could achieve very high accuracy in keyword spotting if we were able to realize perfect phoneme recognition.
Unfortunately, phoneme recognition itself is an immature technology. 
It requires large quantities of labeled data to train the phone language model, and
most of existing phoneme recognizers produce undesirable results due to the flaw of high error rate or inefficiency \cite{cmu-phone-recog,weber2016progress}.
The phone error rate ranges from $30\%-60\%$ but researchers are still making efforts to improve the accuracy \cite{weber2016progress}.
We believe with a better phoneme recognizer our \vs can hopefully achieve better accuracy and efficiency in the future.

\textbf{Sensitive speech content understanding:}
Text-based content privacy preservation has always been a difficult problem in the literature \cite{jansen2006search}. 
The major reason is that there is no explicit definition for content privacy,
as it is context-specific and privacy itself is user-specific.
For instance, the word ``doctor'' might be sensitive in the context ``I have to see my doctor ASAP'' but less sensitive 
in another context like ``I dreamed to be a doctor when I was young''. Also, some people think seeing a doctor is a private affair for them but for others it doesn't matter to let strangers know that.
Protecting content privacy with voice  is even  more challenging,
as transforming speech to text  is computationally expensive and inaccurate by  mobile devices.
}

\textbf{Speech segmentation and randomization:}
There is an alternative to prevent the adversary from performing voice conversion on the speech reversely.
The \vs executes speech segmentation and randomizes voice conversion on each segment.
We observe by experiment that if a word or phrase is split into two halves and they are transformed to two different voices, they can hardly be recognized by the cloud. 
We can \emph{randomly} partition the speech into segments with random size (segments are separated by the gaps between words) and then 
independently convert the voice of each segment with randomly selected parameters.

\textbf{Compatibility with voice assistants:}
We aimed to protect the identity privacy of users of voice input and voice search. 
For voice assistant,  some services activated by voice commands require authentication of the user (by other means, \eg password, two-factor), \eg creating an event on the calendar. The user's PII is inevitably exposed to the voice assistant, which appears to be against our goal.
Today's voice assistants like Google Assistant usually works by three steps: \sr (voice input), command understanding, and command execution.
The first step is not necessary for a voice assistant. We suggest it be decoupled from the voice assistant because users experience a higher privacy leak risk when all their data are stored at one single party. Instead, the voice input task can be taken on by a third-party dedicated to \sr and users connect to it through the \vs.

\textbf{Assumption about PIIs:} This paper assumed the cloud doesn't have users' PII, which can be achieved by existing widely deployed anonymous networks such as Tor \cite{tor,I2P}. Yet at present, most of the \sr service providers have users' PII.
However, our \vs still helps alleviate the risk of privacy breach.
For one thing, every user's voice is converted to many different voices, which can reduce the chance that the user is recognized in real life. 
For another thing, the \vs hides the sensitive phrases in the speeches so it will be more difficult for the cloud to learn the speech content of the user.

\jnl{
\textbf{Incorporation into Android:}
Another concern is that there are other apps on the smart phone that may record users' voice and harm their privacy. For example, 
Facebook needs access to the microphone for voice to text type posts/comments. There are two possible solutions.
First, now that we have designed \vs to be a voice input keyboard app in this paper, users can simply  
deny the microphone permission request of Facebook and avoid using its voice to text feature.
As an alternative, if such apps forces users to provide the microphone permission but they have to use these apps, we can incorporate the \vs into 
Android so that all apps must access it for audio signals. The \vs perturbs the raw audio only when it detects human speaking.
}

\section{Related Work}
\label{sec:related}

Security and privacy on voice data have been concentrated on in previous works from two main aspects.
Spoofing attack to speaker verification systems has received considerable attention in the area of speaker recognition. 
The simplest methods are replay and human impersonation.
More powerful techniques include speech synthesis  \cite{de2012evaluation} and voice conversion \cite{wu2014voice}.
Meanwhile, many robust speaker verification techniques have been proposed to defend against those spoofing attacks 
\cite{wu2014study,wu2015spoofing}. 
Other researchers \cite{pathak2012privacy,pathak2013privacy} proposed to use secure multi-party computation (SMC) 
to implement and achieve speaker verification in a privacy-preserving fashion.
Smaragdis \etal \cite{smaragdis2007framework} were the first to design an SMC-based secure speech recognition, though it is said to be a rudimentary system and the cryptographic protocols are computationally expensive.
In addition, privacy learning has also been a focus in spoken language analysis.
For example, Dan Gillick \cite{gillick2010can} showed that word use preference can be
utilized to infer the speaker's demographics including gender, age, ethnicity, birth places and social-status.
Mairesse \etal \cite{mairesse2007using} designed classification, regression and ranking models to 
learn the Big Five personality traits of the speaker including ``extraversion vs. introversion''.
Other sensitive information contained in the utterance such as emotions \cite{patel2017emotion} and 
health state \cite{schuller2011interspeech} could also be inferred by speaker classification methods. 

Another related work by us \cite{qian2018towards} studied and quantified the privacy risks in speech data publishing and proposed privacy-preserving countermeasures.
By the way, there is a body of work on distributed \sr \eg \cite{zhang2000study} for mobile devices. 
It enables the device itself to extract spectral features from the speech and then send them to the cloud where the features are converted to text. 
Distributed \sr is not commonly used yet and the main purpose is to reduce the communication cost.
Though the user's voice is not directly exposed,  still the features like MFCC can be used to extract voice print and reconstruct the voice accurately.

\section{Conclusion}
\label{sec:conclusion}

Now we are speeding toward a not-too-distant future when we can perform human-computer 
interaction using solely our voice.
In this work, we present a light-weight voice sanitization app that achieves a good protection of both user identity and private speech content, with minimal degradation in the quality of speech recognition.
Our system adopts a carefully designed voice conversion mechanism that is robust to several attacks.
Meanwhile, it utilizes an evolution-based keyword substitution technique to sanitize the voice input content.
Possible future work includes explicitly defining identity privacy for voice input users and achieving adaptive speech recognition while preserving identity privacy.

\appendices
\section{Proof of Theorem 1} 
\label{app1}

\begin{IEEEproof}
This proof is inspired by \cite{erlingsson2014rappor}.
Let $B'=[b'_1,b'_2]$ be the new Bloom filter we get after Step 3 of PRAKA.
For any $i=1,2$, we have
$$P(b'_i=b_i \mid b_i)=1-p+\dfrac{p}{2}=1-\dfrac{p}{2},$$
$$P(b'_i\neq b_i \mid b_i)=\dfrac{p}{2}.$$
For any two distinct values of $B$,  say $B_1=[0,1]$ and $B_2=[1,0]$ without loss of generality,
\begin{equation}
\begin{aligned}
& \dfrac{P(B'\in R \mid B=B_1)}{P(B'\in R \mid B=B_2)} \notag \\
=& \dfrac{\sum_{B'\in R}P(B'\mid B=B_1)}{\sum_{B_k\in R}P(B'\mid B=B_2)} \notag \\
\leq & \max_{B'\in R}\dfrac{P(B'\mid B=B_1)}{P(B'\mid B=B_2)} \notag \\
=& \max_{b'_1,b'_2}\dfrac{\left(\dfrac{p}{2}\right)^{b'_1}\left(1-\dfrac{p}{2}\right)^{1-b'_1} \cdot \left(1-\dfrac{p}{2}\right)^{b'_2}\left(\dfrac{p}{2}\right)^{1-b'_2}}  {\left(1-\dfrac{p}{2}\right)^{b'_1}\left(\dfrac{p}{2}\right)^{1-b'_1} \cdot \left(\dfrac{p}{2}\right)^{b'_2}\left(1-\dfrac{p}{2}\right)^{1-b'_2}} \notag \\
=& \max_{b'_1,b'_2}\left(\dfrac{p}{2}\right)^{2(b'_1-b'_2)}\left(1-\dfrac{p}{2}\right)^{2(b'_2-b'_1)} \notag \\
=& \max_{b'_1,b'_2}\left(\dfrac{2-p}{p}\right)^{2(b'_2-b'_1)} \notag \\
\leq & \left(\dfrac{2-p}{p}\right)^2 (since 0\leq p\leq 1).
\end{aligned}
\end{equation}
Hence, PRAKA satisfies $\epsilon$-DP where $\epsilon=2\ln \frac{2-p}{p}$.
\end{IEEEproof}

\section{Error Bound of $\hat{n}$} 
\label{app2}
Let $n_0$ be the true number of users who set $b_1=1$, 
then $n=X+Y$ where $X\sim B(N-n_0,\frac{p}{2}), Y\sim B(n_0,1-\frac{p}{2}).$
The probability mass function of $n$ is
\begin{equation}
\small
\begin{aligned}
& Pr(n=k) \\
& =\sum_{i=0}^{k}\binom{N-n_0}{i}\binom{n_0}{k-i} \left(\frac{p}{2}\right)^{n_0-k+2i} \left(1-\frac{p}{2}\right)^{N-n_0+k-2i}.
\end{aligned}
\end{equation}

Since $\hat{n}=(n-\frac{p}{2}N)/(1-p)$, we have the error bound
\begin{equation}
\small
\begin{aligned}
& Pr(|\hat{n}-n_0|\le\epsilon) \\
& =Pr(-\epsilon\le\hat{n}-n_0\le\epsilon) \\
& = Pr((1-p)(n_0-\epsilon)+\frac{p}{2}N\le n\le (1-p)(n_0+\epsilon)+\frac{p}{2}N) \\
& = \sum_{k=\lceil (1-p)(n_0-\epsilon)+\frac{p}{2}N\rceil}^{\lfloor (1-p)(n_0+\epsilon)+\frac{p}{2}N\rfloor} Pr(n=k).
\end{aligned}
\end{equation}

\bibliographystyle{abbrv}
\bibliography{voice-sanitizer-v2}

\end{document}